\documentclass[sigconf,natbib=true]{acmart}
\AtBeginDocument{%
  \providecommand\BibTeX{{%
    \normalfont B\kern-0.5em{\scshape i\kern-0.25em b}\kern-0.8em\TeX}}}

\usepackage{array}
\usepackage{subfigure}
\usepackage{graphicx}
\usepackage[switch]{lineno}
\usepackage{amsmath}
\usepackage{multirow}
\usepackage{booktabs}
\usepackage{bbding}
\usepackage{algorithm}
\usepackage{algorithmic}
\usepackage{xcolor}
\usepackage{xspace}
\usepackage{arydshln}
\usepackage{multicol}

\newcommand{\ie}{\emph{i.e.,}\xspace}

\newcommand{\baby}{{FineRec}\xspace}
\newcommand{\babyx}{{FineRec}}

\newcommand{\fig}{Figure\xspace}
\newtheorem*{template}{Prompt}

\copyrightyear{2024}
\acmYear{2024}
\setcopyright{acmlicensed}\acmConference[SIGIR '24]{Proceedings of the 47th International ACM SIGIR Conference on Research and Development in Information Retrieval}{July 14--18, 2024}{Washington, DC, USA}
\acmBooktitle{Proceedings of the 47th International ACM SIGIR Conference on Research and Development in Information Retrieval (SIGIR '24), July 14--18, 2024, Washington, DC, USA}
\acmDOI{10.1145/3626772.3657761}
\acmISBN{979-8-4007-0431-4/24/07}

\begin{document}

\title{FineRec: Exploring Fine-grained Sequential Recommendation}

\author{Xiaokun Zhang}
\authornote{This work was done during Xiaokun visiting Prof.Fenglong Ma at Peen State.}
\affiliation{%
  \institution{Dalian University of Technology}
  \city{}
  \country{}
}
\email{dawnkun1993@gmail.com}

\author{Bo Xu}
\authornote{Corresponding Author.}
\affiliation{%
  \institution{Dalian University of Technology}
  \city{}
  \country{}
  }
\email{xubo@dlut.edu.cn}

\author{Youlin Wu}
\affiliation{%
  \institution{Dalian University of Technology}
  \city{}
  \country{}
  }
\email{wuyoulin@mail.dlut.edu.cn}

\author{Yuan Zhong}
\affiliation{%
  \institution{Pennsylvania State University}
  \city{}
  \country{}
  }
\email{yfz5556@psu.edu}

\author{Hongfei Lin}
\affiliation{%
  \institution{Dalian University of Technology}
  \city{}
  \country{}
  }
\email{hflin@dlut.edu.cn}

\author{Fenglong Ma}
\affiliation{%
  \institution{Pennsylvania State University}
  \city{}
  \country{}
  }
\email{fenglong@psu.edu}

\renewcommand{\shortauthors}{Xiaokun Zhang, et al.}

\begin{abstract}
Sequential recommendation is dedicated to offering items of interest for users based on their history behaviors. The attribute-opinion pairs, expressed by users in their reviews for items, provide the potentials to capture user preferences and item characteristics at a fine-grained level. To this end, we propose a novel framework \baby that explores the attribute-opinion pairs of reviews to finely handle sequential recommendation. Specifically, we utilize a large language model to extract attribute-opinion pairs from reviews. For each attribute, a unique attribute-specific user-opinion-item graph is created, where corresponding opinions serve as the edges linking heterogeneous user and item nodes. To tackle the diversity of opinions, we devise a diversity-aware convolution operation to aggregate information within the graphs, enabling attribute-specific user and item representation learning. Ultimately, we present an interaction-driven fusion mechanism to integrate attribute-specific user/item representations across all attributes for generating recommendations. Extensive experiments conducted on several real-world datasets demonstrate the superiority of our \baby over existing state-of-the-art methods. Further analysis also verifies the effectiveness of our fine-grained manner in handling the task.
\end{abstract}





\maketitle

\section{Introduction}
Sequential Recommendation (SR) aims to provide items of interest for users based on their history behaviors~\cite{GRU4Rec, SASRec}. SR concentrates on addressing the inherent feature of user behaviors, \ie chronologically evolving preferences. This focus equips SR with the ability to grasp users' shifting interests, deliver personalized services timely, and adapt to the fast-paced trends of the modern information age. Owing to such merits, SR has received widespread attention from both academia and industry in recent years~\cite{Li@KDD2023, Du@SIGIR2023, Yang@SIGIR2023, zhang2024side}. 

\begin{figure}[t]
  \centering
  \includegraphics[width=0.9\linewidth]{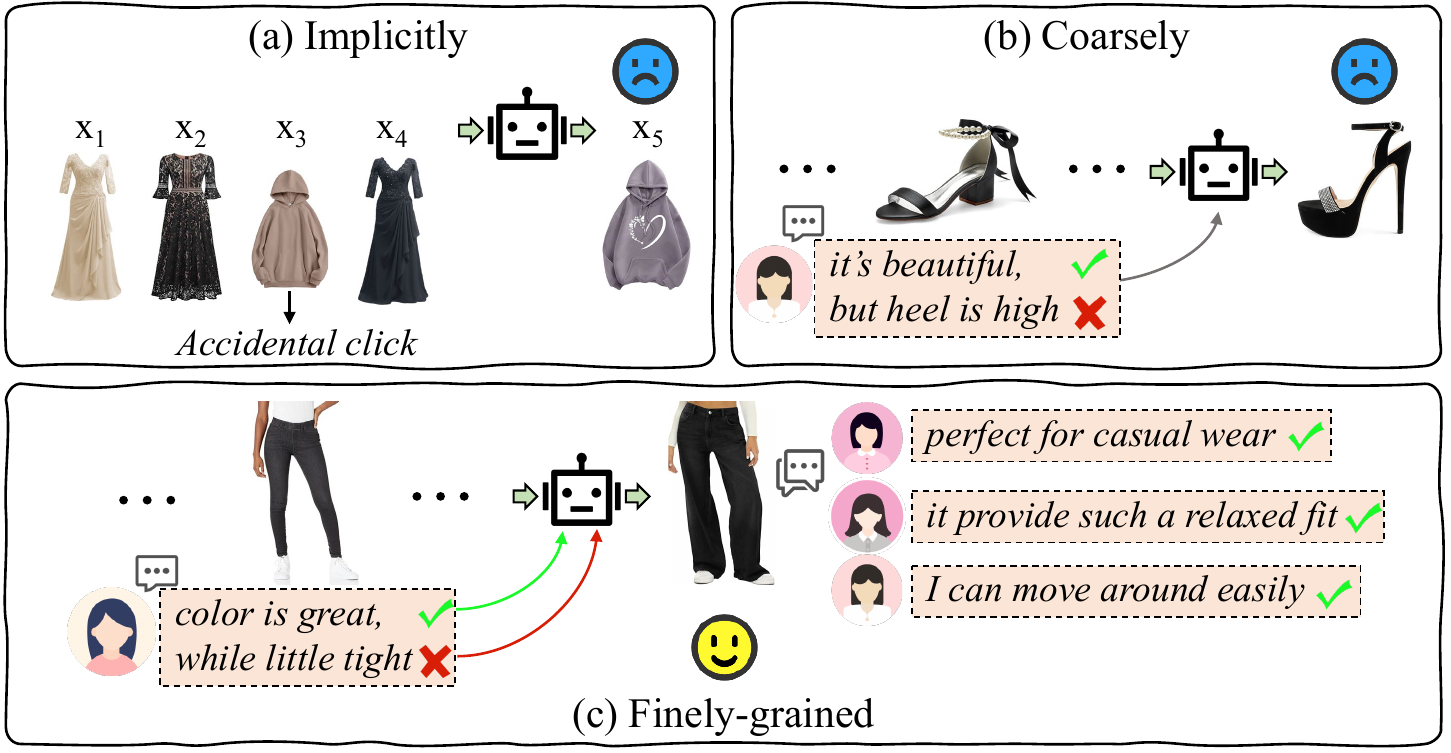}
  \caption{The limitations of existing implicit (a) and coarse (b) recommendations. Our proposed fine-grained manner(c).}\label{intro}
\vspace{-0.1in}
\end{figure}

Existing methods typically rely on neural networks to obtain item and user embeddings, basing recommendations on the similarity between these embeddings. Most efforts attend to model \textbf{implicit} activities of users, like clicks, to infer their preferences, employing various neural architectures including Recurrent Neural Networks (RNN)~\cite{GRU4Rec, NARM}, attention mechanism~\cite{SASRec, BERT4Rec} and Graph Neural Networks (GNN)~\cite{SR-GNN, Ye@SIGIR2023}. 
As shown in \fig~\ref{intro} (a), unfortunately, a significant drawback of this implicit approach is that implicit actions, like accidental clicks, can sometimes misrepresent user interests, thereby introducing noise into the models~\cite{Zhang@KDD2023, Ye@SIGIR2023}.
% As shown in \fig~\ref{intro} (a), unfortunately, a user's implicit action can not convey her interest in some cases, where a mis-click provides nothing but noise for the models. Thus, the implicit manner of these methods has inherent limitations. 
Another line of research~\cite{RNS, Dong@AAAI2020, Shuai@SIGIR2022} utilizes user-item reviews to explicitly capture user preferences. However, they all \textbf{coarsely} treat a review as a whole, overlooking the fact that users may have distinct opinions about different attributes within a single review. As presented in \fig~\ref{intro} (b), this failure to discern such fine-grained distinctions limits the effectiveness of these methods.
% a user may hold entirely different attitudes towards certain attributes of an item. Indiscriminately blending various opinions in a review limits the effectiveness of such methods.

Actually, in a review, a user expresses her specific feelings for an item in the form of \textbf{attribute-opinion} pairs. These pairs provide an opportunity to capture user preferences and item characteristics at an \textbf{explicit and fine-grained} manner, offering a new perspective to improve recommendation. As depicted in \fig~\ref{intro} (c), this manner facilitates satisfactory recommendations, such as suggesting loose clothing to a user who previously expressed a dislike for tight fits, and the new item is favorably described as loose by other users. Despite holding encouraging prospects, effectively implementing the fine-grained manner in SR faces several obstacles:
% in such a manner, we can accurately recommend a new clothing with loose style to meet a user's need, where the user expressed her dislike for tight clothing before and the new one is rated as loose by other users. Despite holding encouraging prospects, unfortunately, it is non-trivial to finely handle sequential recommendation in the attribute-opinion manner due to following obstacles:
% As presented in \fig~\ref{intro} (c), insights into a user's likes and dislikes for specific attributes can be derived from the attribute-opinion pairs.

Firstly, it is difficult to \emph{extract informative attribute-opinion pairs} from reviews. Users often express their opinions using informal and implicit language, which complicates the extraction. For instance, consider the review, "I was robbed by this thing!"--neither the attribute (Price) nor opinion (Expensive) is delivered in a straightforward way. Lacking rich language knowledge, current methods, either based on hand-craft rules~\cite{Hu@KDD2004} or well-designed models~\cite{Mohna@KDD2022} tailored for certain fields, can not accurately extract informative attribute-opinion pairs from such reviews. 

Secondly, a major challenge lies in \emph{finely representing users and items} under each attribute with diverse opinions. On one hand, users and items exhibit unique preferences and characteristics under different attributes. However, current methods typically represent them in a overall way, overlooking the distinctions across attributes. On the other hand, even in a certain attribute, the opinion diversity impedes the learning of fine-grained user and item representations. For users, their attitudes would be changing along with distinct items on an attribute. For example, a user may prefer televisions with large size for enhanced viewing, while favoring small-sized phones due to portability. As to items, they may receive various ratings from different users under an attribute. It is common that some users favor brightly colored clothes, while others may hate them. Thus, it is imperative to consider opinion diversity to achieve fine-grained user and item representations at attribute-level.

Thirdly, it remains a problem to \emph{generate recommendations} in the fine-grained manner. User behaviors are determined by user preferences and item characteristics in various attributes jointly. For instance, a loyal fan may buy an item of her favored brand despite disliking its high price. Unfortunately, we can only observe the overall user-item interactions, while there are no explicit indicators to show how each attribute affects user decisions. This makes it challenging to infer user actions in the fine-grained manner.

To tackle these issues, we propose a novel framework to incorporate attribute-opinion pairs for \underline{Fine}ly handling sequential \underline{Rec}ommendation (\baby). Firstly, trained on massive text data, the Large Language Model (LLM) encapsulates a wealth of language knowledge~\cite{Bao@RecSys2023, Liu@CIKM2023Text, Geng@EMNLP2023}. In light of this, we utilize LLM to extract attribute-opinion pairs from reviews. To relieve hallucinations of LLMs in tackling complex tasks, we obtain attributes based on their importance on websites and subsequently extract corresponding opinions via LLM. Secondly, to represent users/items under distinct attributes, we create a unique attribute-specific user-opinion-item graph for each attribute. Within the graph, heterogeneous user and item nodes are connected by opinion edges which represent the specific opinions users have expressed about items under the attribute. Afterwards, a diversity-aware convolution operation is devised to achieve fine-grained user and item representation on the graphs, with an emphasis on opinion diversity during the information aggregation. Lastly, we present an interaction-driven fusion mechanism that leverages user-item interaction information to guide the fusion of attribute-specific item/user representations, achieving final recommendations.
In summary, the contributions of this study are outlined as follows:
% Finally, \baby generates recommendations based on comprehensive item/user representations derived from various attribute-opinions.
\begin{itemize}
    \item We present a new perspective to handle sequential recommendation, where the attribute-opinions of reviews are explored to finely reveal user preferences and item characteristics. As far as we know, this is the pioneering attempt to handle the task in such a fine-grained manner. 
    \item We proposed a novel framework \baby comprising several innovative techniques, including LLM-based attribute-opinion extraction, attribute-specific user-opinion-item graph with a diversity-aware convolution operation, and an interaction-driven fusion mechanism, to achieve fine-grained sequential recommendation.
    \item Extensive experiments on several public benchmarks validate the superiority of our \baby over state-of-the-art methods. Further analysis also justifies the effectiveness of our fine-grained manner in handling the task.
\end{itemize}

\section{Related Work}

\subsection{Sequential Recommendation}
With the powerful ability in representing users and items, neural networks have come to dominate the field of sequential recommendation~\cite{Du@SIGIR2023, Liu@CIKM2023}. Various neural structures are employed to model user behavior sequences like RNN~\cite{GRU4Rec, NARM}, attention mechanism~\cite{SASRec, DIDN, Zhang@WSDM2023}, Transformers~\cite{BERT4Rec, Zhang@KDD2023, Zhou@CIKM2023}, GNN~\cite{SR-GNN, Ye@SIGIR2023}, MLP~\cite{li@IJCAI2022, Li@WWW2023} and contrastive learning~\cite{chen@WWW2022, Qin@SIGIR2023}. Some recent methods incorporate extra item information to mine user preferences from multiple views including categories~\cite{cai@SIGIR2021, Liu@RecSys2023, Chen@SIGIR2023}, brands~\cite{song@CIKM2021, Du@SIGIR2023}, price~\cite{CoHHN, BiPNet}, text~\cite{hou@KDD2022, Li@KDD2023, Liu@CIKM2023Text} as well as images~\cite{Hu@CIKM2023, MMSBR}. However, all of them focus on modeling user implicit behaviors which can not represent genuine intents of users in some cases. Thus, this limitation significantly hampers their effectiveness.

\subsection{Review-driven Recommendation}
Reviews, posted by users to explicitly convey their opinions on items, are utilized by recent works to improve recommendation task~\cite{RNS, Dong@AAAI2020, Shuai@SIGIR2022}. These works typically employ reviews to build user and item embeddings, and then rely on certain neural structures to offer suggestions like attention mechanism~\cite{Liu@KDD2019, Dong@AAAI2020} and GNN~\cite{Shuai@SIGIR2022}. RNS~\cite{RNS} is a representative work incorporating reviews into sequential recommendation, where review contents are utilized to obtain users' long-term and short-term preferences. However, these efforts ignore the significant differences in a user review regarding different attributes, and instead coarsely model a review as a whole. 
% Zhang@SIGIR2014
Moreover, some methods attempt to capture user distinct interest on varying aspects~\cite{Zhang@SIGIR2014, Cheng@IJCAI2018, Chin@CIKM2018, Ren@IJCAI2023}. Unfortunately, they only focus on discerning various aspects while do not capture specific user opinions towards these aspects. Such paradigms fall short of finely disclosing user preferences or item characteristics either.

\begin{figure*}[t]
  \centering
  \includegraphics[width=0.90\linewidth]{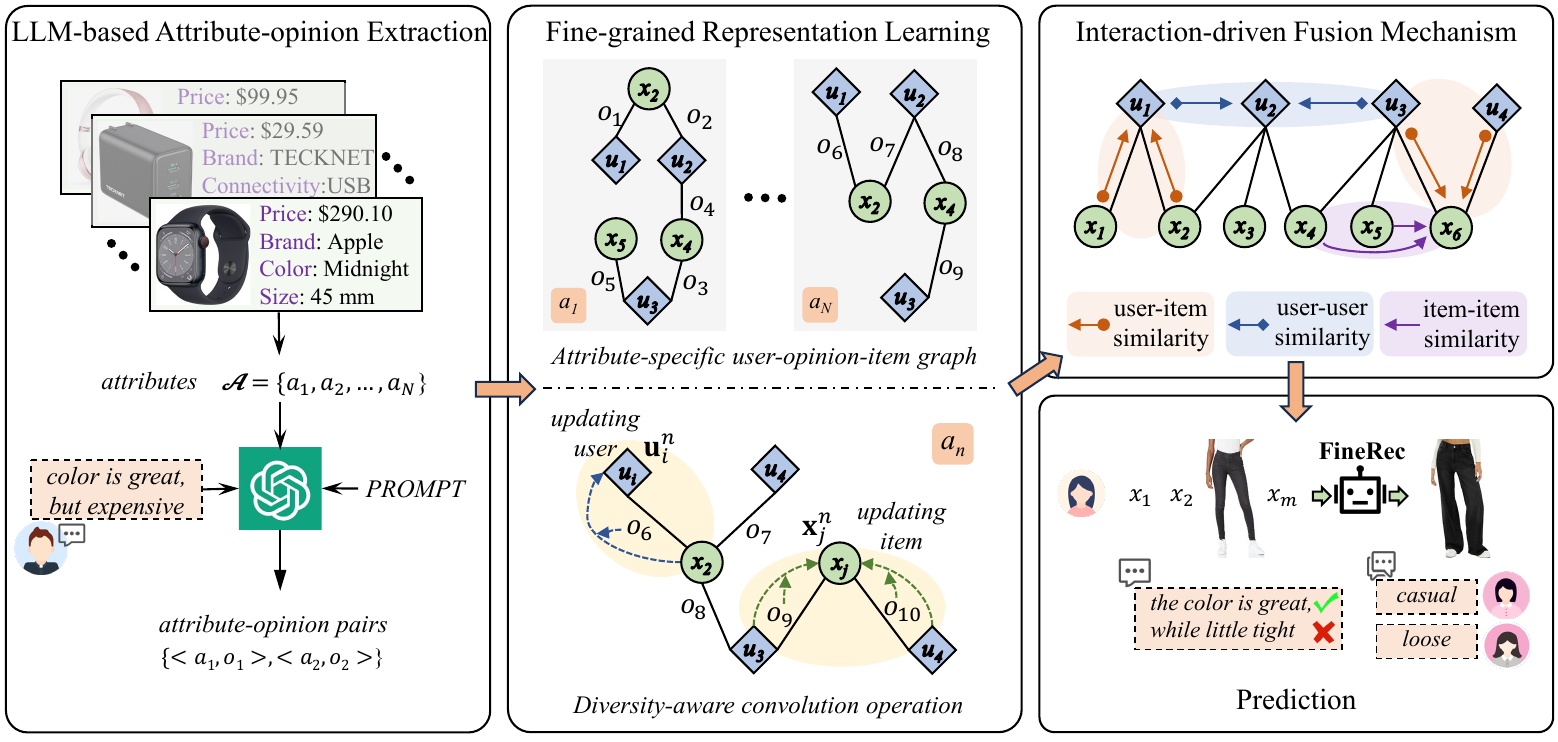}
  \caption{The workflow of \baby. We employ LLM to extract attribute-opinion pairs from user-item reviews. Afterwards, an attribute-specific user-opinion-item graph is created in each attribute, where a diversity-aware convolution operation is devised to conduct information aggregating. Interaction-driven fusion mechanism leverages user-item interactions to integrate attribute-specific item/user representations. Finally, \baby generates recommendations in the fine-grained manner.}\label{finerec}
\end{figure*}

\section{Problem Statement}
% The goal of sequential recommendation is to predict users' next interacted items based on their previous behavior sequences. 
Let $\mathcal{U}$ and $\mathcal{X}$ denote the unique user and item set, respectively. The $\mathcal{R}$ is a review set consisting of all reviews posted by users on items. Each user $u_i \in \mathcal{U}$ has interacted with a sequence of items chronologically, producing a behavior sequence denoted as: $\mathcal{S}^{u_i}$ = [$x_1, x_2, \cdots, x_m$], where $x_j$ $\in$ $\mathcal{X}$ and $m$ is the sequence length. The review posted by user $u_i$ on item $x_j$ is denoted as $r_{ij} \in \mathcal{R}$. In a review, the attribute-opinion pairs depict user preferences and item characteristics in a fine-grained manner. In this study, therefore, we propose to finely represent users and items via exploring the attribute-opinion pairs from reviews. Accordingly, we can forecast the user next interacted item $x_{m+1}$ based on $\mathcal{S}^{u_i}$.

\section{The Proposed Approach}

\subsection{Overview of the \baby}
As illustrated in \fig~\ref{finerec}, the proposed \baby mainly consists of the following interconnected components: (1) \textbf{LLM-based attribute-opinion extraction} identifies informative attributes based on their importance and subsequently extracts the attribute-opinion pairs from reviews via LLM; (2) \textbf{Fine-grained representation learning} obtains fine-grained user and item embeddings via exploring attribute-opinion pairs, where \emph{an attribute-specific user-opinion-item graph} is created under each attribute, with \emph{a diversity-aware convolution operation} conducting information aggregating; (3) \textbf{Interaction-driven fusion mechanism} leverages user-item interaction information to guide the fusion of attribute-specific item/user representations; and (4) \textbf{Prediction} provides recommendations in the fine-grained manner.

\subsection{LLM-based Attribute-opinion Extraction}
Large Language Models (LLMs), benefiting from their training on massive text corpora, have become reservoirs of vast language knowledge. This foundation empowers LLMs with remarkable proficiency in various tasks of Natural Language Processing (NLP)~\cite{Bao@RecSys2023, Liu@CIKM2023Text, Geng@EMNLP2023}. Leveraging this capability, we utilize LLMs for extracting attribute-opinion pairs from reviews. In the context of recommendation, the item attributes that users care about are heavily overlapping, that is, a small number of important attributes significantly influence user choice like item price~\cite{CoHHN}. Besides, LLMs can sometimes suffer from "hallucinations" where they perform unsatisfactorily in addressing complex issues~\cite{Min@CSUR2024}. In light of these facts, we present to obtain informative attributes based on their importance, and subsequently extract corresponding opinions via LLMs. Such a paradigm enables \baby to focus on important attributes and greatly simplifies the extraction task.

Specifically, we gather item attributes directly from E-commerce websites, with an assumption that the displayed attributes are important. Adhering to the Pareto principle, we focus on the most prevalent attributes, selecting the top $N$ attributes based on their frequency of occurrence to build the attribute set $\mathcal{A}$ = \{$ a_1, a_2, \cdots, a_N$\}, where $N$ is the number of extracted attributes. Note that, each $a_i$ represents an informative attribute like `Price', `Color' and so on. 

A well-crafted prompt is the key to unleash the potential of LLMs. Building upon the extracted attributes, thus, we develop a specialized prompt template to facilitate the LLM in extracting attribute-opinion pairs from reviews as follows,

\begin{template}
    Please extract user opinion words towards the attribute "$a_n$" from the review: "review contents". Only return the opinion words! Your answer should be short.
\end{template}
where $a_n \in \mathcal{A}$ and \emph{"review contents"} denotes a specific review. For a review $r_{ij}$, the LLM outputs a sentence that captures the opinion of a user ($u_i$) about an item ($x_j$) under the attribute ($a_n$), denoted as $o^{n}_{ij} \in \mathcal{O}$, where $\mathcal{O}$ is opinion set. If the review does not mention any attribute in $\mathcal{A}$, the result is disregarded. Note that, we apply LLM to conduct the extraction across all reviews for every attribute, forming the attribute-opinion pairs. For instance, consider a review $r_{ij} \in \mathcal{R}$ from user $u_i$ about item $x_j$ stating, "\emph{it smells nice, but too expensive}", with the attribute set $\mathcal{A}$ = \{$Scent, Price, Brand$\}, the extracted attribute-opinion pairs would be \{ <Scent, nice>, <Price, expensive> \}. These extracted pairs allow us to finely represent users and items based on their specific attributes and opinions.

\subsection{Fine-grained Representation Learning}
\subsubsection{Attribute-specific user-opinion-item graph}

Generally, a/an user/item holds distinct preferences/characteristics under different attributes. Therefore, we create an attribute-specific user-opinion-item graph to finely-grained encode users/items under each attribute. 
% A toy example is presented in the \fig~\ref{finerec}. 
Specifically, for an attribute $a_n$, a unique user-opinion-item graph $\mathcal{G}^n$ = $(\mathcal{U}^n \bigcup \mathcal{X}^n, \mathcal{O}^n)$ is built. In this graph, $\mathcal{U}^n \bigcup \mathcal{X}^n$ is the node set consisting of users and items, where $\mathcal{U}^n  \subseteq \mathcal{U}$ and $ \mathcal{X}^n \subseteq \mathcal{X}$. The $\mathcal{O}^n \subseteq \mathcal{O}$ is the edge set encapsulating opinions expressed by users about items regarding the specific attribute. As shown in \fig~\ref{finerec}, within the graph $\mathcal{G}^n$ of the attribute $a_n$, a user $u_i \in \mathcal{U}^n$ and an item $ x_j \in \mathcal{X}^n$ are connected by an opinion edge $o_{ij}^{n} \in \mathcal{O}^n$ if the user expressed her opinion $o^n_{ij}$ to the item in relation to the attribute. Note that, the $\mathcal{G}^n$ will not record the user and item if there is not an opinion between them under the attribute. Moreover, the attribute-specific user-opinion-item graph captures interaction and opinion information between users and items, providing the potentials to discern fine-grained user preferences and item characteristics specific to each attribute.

% We initiate user and item representations under distinct attributes with different look-up embedding tables. 
To represent various user/item preferences/characteristics under different attributes, we employ distinct embeddings to represent users/items in each attribute.
To elaborate, a user $u_i$ is separately represented by $N$ unique embeddings as \{$\mathbf{u}_i^{1}, \mathbf{u}_i^{2}, \cdots, \mathbf{u}_i^{N}$\}, where $\mathbf{u}_i^{n} \in \mathbb{R}^d$ indicates the user preferences about attribute $a_n \in \mathcal{A}$. The same applies to items, \ie each item $x_j$ is represented as \{$\mathbf{x}_j^{1}, \mathbf{x}_j^{2}, \cdots, \mathbf{x}_j^{N}$\}, where $\mathbf{x}_j^{n} \in \mathbb{R}^d$ denotes the item characteristics in attribute $a_n$.
Besides, we employ pre-trained BERT~\cite{BERT} to represent an attribute $a_n$ as $\mathbf{a}^{n} \in \mathbb{R}^d$ and  an opinion text $o^{n}_{ij}$ as $\mathbf{o}_{ij}^{n} \in \mathbb{R}^d$. 

\subsubsection{Diversity-aware convolution operation}
The extensive diversity of opinions impedes fine-grained user and item representation learning through attribute-opinion manner. Specifically, for a given attribute, a user may have varying opinions about different items, and similarly, an item might receive a range of opinions from different users. To address this diversity, we devise a diversity-aware convolution operation to conduct information aggregating on each attribute-specific user-opinion-item graph. Formally, we update a user's embedding $\mathbf{u}_{i}^{n}$ on the $\mathcal{G}^n$ of attribute $\mathbf{a}^{n}$ via,
\begin{align}
    % \mathbf{e}_{i}^{u:n} &= f(\mathbf{e}_{i}^{u:n}, \mathbf{e}^{a:n}, I^{u:n}, \mathcal{O}^{u:n})
    \mathbf{u}_{i}^{n} &:= \mathbf{u}_{i}^{n} + \sum_{{x}_j^{n} \in \mathcal{X}^n_{u_i}} \alpha_j (\mathbf{x}_j^{n} + \mathbf{o}_{ij}^{n})
\end{align}
where $\mathcal{X}^n_{u_i} \subseteq \mathcal{X}^n$ consists of items adjacent with the user in the graph $\mathcal{G}^n$, while $\mathbf{o}^{n}_{ij}$ is embedding of corresponding opinion. It's important to highlight that our method updates user embeddings by jointly considering both the item and its associated opinion (instantiated with an additive operation). This approach allows our \baby to capture the diversity of a user's opinions towards different items. The importance of different item-opinion pairs is determined by,
\begin{align}
    % \mathbf{e}_{i}^{u:n} &= f(\mathbf{e}_{i}^{u:n}, \mathbf{e}^{a:n}, I^{u:n}, \mathcal{O}^{u:n})
    \alpha_j &= \frac{sim(\mathbf{u}_{i}^{n}, (\mathbf{a}^{n}+\mathbf{x}_j^{n}))}{\sum_{x_k^n \in \mathcal{X}^n_{u_i}}sim(\mathbf{u}_{i}^{n}, (\mathbf{a}^{n}+\mathbf{x}_k^{n}))},
\end{align}
where the $sim()$ is cosine similarity. To highlight the attribute's influence on user behaviors, we employ the attribute embedding $\mathbf{a}^{n}$ to calculate the importance. 
Similarly, we update an item embedding $\mathbf{x}_{j}^{n}$ on the attribute-specific graph $\mathcal{G}^n$ as follows, 
\begin{align}
    \mathbf{x}_{j}^{n} &:= \mathbf{x}_{j}^{n} + \sum_{u_i^{n} \in \mathcal{U}^n_{x_j}} \beta_i (\mathbf{u}_i^{n} + \mathbf{o}_{ij}^{n}) \\
    \beta_i &= \frac{sim(\mathbf{x}_{j}^{n}, (\mathbf{a}^{n}+\mathbf{u}_i^{n}))}{\sum_{u_k^n \in \mathcal{U}^n_{x_j}}sim(\mathbf{x}_{j}^{n}, (\mathbf{a}^{n}+\mathbf{u}_k^{n}))}
\end{align}
where $\mathcal{U}^n_{x_j}\subseteq\mathcal{U}^n$ contains users expressing opinions on the item under the attribute $a_n$.

\subsection{Interaction-driven Fusion Mechanism}
The future actions of users are jointly influenced by their preferences and items' characteristics across various attributes. 
% Consequently, it is essential to integrate user and item embeddings from different attributes for accurate behavior prediction. 
However, capturing the intricate patterns of the influence is challenging, primarily because there are no explicit signals indicating how each attribute impacts user decisions. Fortunately, the user-item interactions provide insights into these complex influencing patterns. For example, when a user selects certain items from a vast choices, it reflects a match between her overall preferences and the characteristics of those items. In fact, these interactions hint at the similarity at the embedding level, considering neural models operate based on the similarity of representations. Thus, we propose to employ the similarity implied in the user-item interactions to integrate attribute-specific item/user representations. 

To intuitively illustrate user-item interaction relations, we present a global user-item interaction graph in the upper right part of~\fig~\ref{finerec}.
As shown in the figure, user-item interactions manifest three types of similarities: (1) user-item similarity, indicating the commonality between a user and her interacted items; (2) user-user similarity, where users buying the same item can be considered similar; (3) item-item similarity, where items interacted with the same user are deemed similar. We start by concatenating attribute-specific item and user embeddings as,
\begin{align}
    \mathbf{\hat{u}}_i &= [\mathbf{u}_{i}^{1}; \mathbf{u}_{i}^{2}; \cdots; \mathbf{u}_{i}^{N}] \\
    \mathbf{\hat{x}}_j &= [\mathbf{x}_{j}^{1}; \mathbf{x}_{j}^{2}; \cdots; \mathbf{x}_{j}^{N}]
\end{align}
where $[;]$ denotes concatenation operation, and $\mathbf{\hat{u}}_i$, $\mathbf{\hat{x}}_j \in \mathbb{R}^{Nd}$. 

We find that, interestingly, three types of similarity can be conceptualized as adjacent relationships within the global user-item graph. Specifically, the user-item similarity represents direct, one-hop connections in the graph, while user-user and item-item similarities correspond to two-hop relationships. In light of aggregating algorithms in GNN, we integrate various preferences of the user ($u_i$) on different attributes as,
\begin{align}
    % \mathbf{e}_i^{u \cdot UI} &= \sum_{\mathbf{\hat{e}}_k^{I} \in \mathcal{I}^{u}} \mathbf{\hat{e}}_k^{I},
    % \mathbf{e}_j^{I \cdot UI} &= \sum_{\mathbf{\hat{e}}_k^{u} \in \mathcal{U}^{I}} \mathbf{\hat{e}}_k^{u} \\
    % \mathbf{e}_i^{u \cdot UU} &= \sum_{\mathbf{\hat{e}}_p^{u} \in \mathcal{U}^{u}} \mathbf{\hat{e}}_p^{u},
    % \mathbf{e}_j^{I \cdot II} &= \sum_{\mathbf{\hat{e}}_q^{I} \in \mathcal{I}^{I}} \mathbf{\hat{e}}_q^{I},
    % \mathbf{u}_i &= \mathbf{\hat{u}}_i + \sum_{\mathbf{\hat{x}}_j \in \mathcal{X}_{u_i}} W_1\mathbf{\hat{x}}_j + \frac{1}{|\mathcal{U}^{u}|} \sum_{\mathbf{\hat{u}}_p \in \mathcal{U}^{u}} \mathbf{\hat{u}}_p,
    \mathbf{u}_i &= \mathbf{\hat{u}}_i + \frac{1}{|\mathcal{X}_{u_i}|}\sum_{{{x}}_j \in \mathcal{X}_{u_i}} W_1\mathbf{\hat{x}}_j +  \frac{1}{|\mathcal{U}_{u_i}|}\sum_{{{u}}_k \in \mathcal{U}_{u_i}}W_2\mathbf{\hat{u}}_k,
\end{align}
where  $W_1, W_2 \in \mathbb{R}^{Nd \times Nd}$ are trainable matrices enabling attribute-specific embeddings to interact with each other, $\mathcal{X}_{u_i}\subseteq\mathcal{X}$ are items interacted by the user $u_i$, $\mathcal{U}_{u_i}\subseteq\mathcal{U}$ contains users buying the same items as the user $u_i$, and $\mathbf{u}_i$ is the final user representation. Note that, $\mathcal{X}_{u_i}$ indicates user-item similarity, while $\mathcal{U}_{u_i}$ denotes user-user similarity. In Eq.(7), for a user, we inject the information of its similar items and users into its embedding. This method highlights the similarities revealed through user-item interactions, urging similar users and items closer in the embedding space. By doing so, it guides the fusion for representations from various attributes. Similarly, we conduct the fusion for an item ($x_j$) as,

\begin{align}
    % \mathbf{e}_i^{u \cdot UI} &= \sum_{\mathbf{\hat{e}}_k^{I} \in \mathcal{I}^{u}} \mathbf{\hat{e}}_k^{I},
    % \mathbf{e}_j^{I \cdot UI} &= \sum_{\mathbf{\hat{e}}_k^{u} \in \mathcal{U}^{I}} \mathbf{\hat{e}}_k^{u} \\
    % \mathbf{e}_i^{u \cdot UU} &= \sum_{\mathbf{\hat{e}}_p^{u} \in \mathcal{U}^{u}} \mathbf{\hat{e}}_p^{u},
    % \mathbf{e}_j^{I \cdot II} &= \sum_{\mathbf{\hat{e}}_q^{I} \in \mathcal{I}^{I}} \mathbf{\hat{e}}_q^{I},
    % \mathbf{x}_j &= \mathbf{\hat{x}}_j + \sum_{\mathbf{\hat{u}}_k \in \mathcal{U}^{I}} W_3 \mathbf{\hat{u}}_k + \frac{1}{|\mathcal{I}^{I}|} \sum_{\mathbf{\hat{v}}_q \in \mathcal{I}^{I}} \mathbf{\hat{v}}_q
    \mathbf{x}_j &= \mathbf{\hat{x}}_j +  \frac{1}{|\mathcal{U}_{x_j}|}\sum_{{{u}}_i \in \mathcal{U}_{x_j}} W_3 \mathbf{\hat{u}}_i + \frac{1}{|\mathcal{X}_{x_j}|}\sum_{{{x}}_k \in \mathcal{X}_{x_j}} W_4\mathbf{\hat{x}}_k,
\end{align}
where $\mathcal{U}_{x_j}\subseteq\mathcal{U}$ denotes users interacting with the item $x_j$, $\mathcal{X}_{x_j}\subseteq\mathcal{X}$ consists of items co-interacted by users with item $x_j$, and $\mathbf{x}_j$ is the final item representation. Besides, $\mathcal{U}_{x_j}$ indicates user-item similarity, while $\mathcal{X}_{x_j}$ denotes item-item similarity.

\subsection{Prediction}
Recent actions of a user can reflect her recent interest. Following~\cite{RNS, SR-GNN, Zhang@WSDM2023}, thus, we represent a user's recent interest $\mathbf{\Bar{u}}_i$ by conducting average-pooling on her last $l$ interacted items as,
\begin{align}
    \mathbf{\Bar{u}}_i = \frac{1}{l} \sum_{k=0}^{l-1} \mathbf{x}_{m-k}.
\end{align}

Based on user embedding $\mathbf{u}_i$, her recent interest $\mathbf{\Bar{u}}_i$ and item embedding $\mathbf{x}_j$, we can forecast next behavior of the user as, 
\begin{align}
    y_j &= (\mathbf{u}_i + \mathbf{\Bar{u}}_i)^{\top}  \mathbf{x}_j,
\end{align}
where $y_j$ is the interacted score predicted for a candidate item $x_j$. To ensure a fair comparison, we follow the training paradigm of full ranking on the entail item set as in~\cite{Zhou@CIKM2023, Zhang@RecSys2023}. Formally, the \baby is trained via cross-entropy loss as follows,
\begin{align}
    \mathcal{L}(\mathbf{p}, \mathbf{y}) = - \sum^{|\mathcal{X}|}_{j=1} p_j \log (y_j) + (1-p_j)\log(1-y_j)
\end{align}
where $p_j$ is the ground truth indicating whether the user purchases item $x_j$.

\section{Experimental Setup}
\subsection{Research Questions}

\begin{table}[t]
% \small
\tabcolsep 0.08in 
\centering
\caption{Statistics of datasets.}
\begin{tabular}{ccccc}
\toprule
Datasets      &Cellphones  &Beauty  & Sports & Yelp \\
\midrule
\#item        & 6,208   & 10,176   & 11,017 & 12,391   \\
\#user     & 7,598  & 15,152   & 11,817   & 12,373 \\
\#interaction & 50,140  & 123,148 & 87,594 &  110,313 \\
avg.length    & 6.60     & 8.13     & 7.41    & 8.92 \\
\bottomrule
\end{tabular}

\label{statistics}
\end{table}

We conduct extensive experiments on several real-world datasets to evaluate the proposed \baby and all baselines, with the focus on answering the following research questions: 

\begin{itemize}
    \item \textbf{RQ1}: How does our \baby perform compared with existing state-of-the-art methods? (ref. Section~\ref{sec:overall})
    
    \item \textbf{RQ2}: What is the effect of the fine-grained attribute-opinion manner in handling the task? (ref. Section~\ref{sec:fine})
    
    \item \textbf{RQ3}: Does each proposed technique contribute positively to \baby's performance? (ref. Section~\ref{sec:diver}-\ref{sec:inter})
    
    \item \textbf{RQ4}: What is the influence of the key hyper-parameters on \baby? (ref. Section~\ref{sec:hyper})

    \item \textbf{RQ5}: How well dose the proposed \baby work in the real-world instance?  (ref. Section~\ref{sec:case})

\end{itemize}

\subsection{Datasets and Preprocessing}

To scrutinize the effectiveness of our \baby, we employ the following four popular public datasets in our experiments:
\begin{itemize}
    
    \item \textbf{Cellphones}, \textbf{Beauty} and \textbf{Sports} are three datasets covering different domains in Amazon\footnote{\url{http://jmcauley.ucsd.edu/data/amazon/}}. As widely used benchmarks for the sequential recommendation~\cite{RNS,Chin@CIKM2018}, these datasets contain users' purchasing behavior sequences and corresponding user-item reviews. 
    
    \item \textbf{Yelp} \footnote{\url{https://www.yelp.com/dataset}} containing users' reviews for restaurants is a popular dataset used in the task~\cite{LXW@WSDM2023,Chin@CIKM2018}. As in~\cite{Lin@WWW2023,LXW@WSDM2023}, we retain the transaction records of the year 2019 in this dataset.
    
\end{itemize}

Following~\cite{SASRec, RNS, hou@KDD2022, Qin@SIGIR2023}, we apply the 5-core method to preprocess these datasets, where items and users with less than 5 interactions are filtered out. Besides, to fairly examine the impact of finely-grained modeling, we eliminate reviews that fail to mention the attributes included in the attribute list $\mathcal{A}$. In line with common practices~\cite{Zhou@CIKM2023, Kim@SIGIR2023}, \emph{leave-one-out} operation is used to split these datasets, where the last item in a sequence is used for testing, the penultimate item for validation, and the remaining for training. We present the statistical details of all four datasets in Table~\ref{statistics}.

\subsection{Evaluation Metrics}

\begin{table}[t]
\small
\tabcolsep 0.01in 
\centering
\caption{Used attributes in each dataset.}
\begin{tabular}{cc}
\toprule
Datasets      & Attribute list\\
\midrule
Cellphones  & Battery, Brand, Color, Connectivity, Performance, Price, Size \\
Beauty      & Brand, Color, Effectiveness, Ingredients, Price, Scent, Size \\
Sports      & Brand, Comfort, Functionality, Material, Price, Quality,  Size \\
Yelp        & Ambience, Cleanliness, Food, Location, Parking, Price, Service\\
\bottomrule
\end{tabular}

\label{attribute}
\end{table}

Following existing works~\cite{RNS, Du@SIGIR2023, Qin@SIGIR2023, Zhou@CIKM2023}, we evaluate the performance of all methods with following two metrics:

\begin{itemize}
\item \textbf{Prec}@$k$: Precision measures the proportion of cases where the ground-truth item is within the recommendation list. 
% \item MRR@k: Mean Reciprocal Rank is the average of reciprocal ranks of the ground-truth items among the recommendation lists. 
% The reciprocal rank is manually set to zero if the rank is larger than k. 
\item \textbf{NDCG}@$k$: Normalized Discounted Cumulative Gain considers the rank of the ground truth item among the recommendation list.
% where hits at higher positions get higher scores.
\end{itemize}

\begin{table*}[ht]
% \small
\tabcolsep 0.05in 
%   \centering
    \caption{Experimental results (\%) of different methods on four datasets. Bold scores are the best performance among all methods, while underlined scores are the second best. Improvements of \baby over the best baseline (*) are statistically significant with $t$-test ($p < 0.01$).}
    \renewcommand{\arraystretch}{1.1}
\begin{tabular}{cccccccccccccc}
\toprule 
Datasets                    & Metrics & SKNN  & NARM  & SASRec    & SR-GNN & RNS  &ICLRec & UniSRec   & A-Mixer & MCLRec              & ACTSR  & \baby           & $impro.$   \\
\midrule 
\multirow{4}{*}{Cellphones} & Prec@10 &2.54    &3.23    &3.95    &4.21    &4.88    &3.59    &4.35    &4.04    &5.83    &\underline{6.12}    &$ \bf 7.66^*$    &25.16\% \\
                            & NDCG@10  &1.45    &1.74    &1.80    &2.02    &\underline{3.18}    &1.54    &2.28    &2.03    &2.78    &2.91    &$ \bf 3.87^*$    &21.70\% \\
                            & Prec@20 &4.01    &4.50    &6.11    &6.03    &7.09    &5.27    &6.29    &5.46    &8.47    &\underline{8.84}    &$ \bf 11.89^*$    &34.50\% \\
                            & NDCG@20 &1.88    &2.06    &2.42    &2.53    &\underline{3.79}    &2.05    &2.66    &2.34    &3.38    &3.55    &$ \bf 4.75^*$    &25.33\% \\
                            \midrule 
\multirow{4}{*}{Beauty}    & Prec@10 &1.73    &2.61    &3.97    &2.25    &2.87    &1.97    &3.71    &3.47    &\underline{4.41}    &4.16    &$ \bf 5.71^*$    &29.48\% \\
                            & NDCG@10  &1.01    &1.51    &1.77    &1.11    &1.52    &1.22    &1.87    &1.91    &\underline{2.08}    &2.05    &$ \bf 2.92^*$    &40.38\%  \\
                            & Prec@20 &2.68    &3.89    &6.64    &3.69    &4.63    &3.04    &6.23    &5.17    &\underline{7.93}    &7.23    &$ \bf 9.25^*$    &16.65\% \\
                            & NDCG@20  &1.76    &1.83    &2.40    &1.52    &2.01    &1.99    &2.14    &2.29    &\underline{2.80}    &2.62    &$ \bf 3.77^*$    &34.64\%  \\
                            \midrule 
\multirow{4}{*}{Sports}     & Prec@10 &1.13    &1.48    &1.69    &1.88    &2.44    &1.59    &1.85    &1.44    &\underline{2.85}    &2.57    &$ \bf 3.50^*$    &22.81\% \\
                            & NDCG@10 &0.73    &0.81    &0.94    &0.91    &\underline{1.27}    &0.90    &0.93    &0.88    &1.17    &1.02    &$ \bf 1.88^*$    &48.03\%  \\
                            & Prec@20 &1.68    &1.84    &2.63    &2.90    &3.11    &2.31    &2.49    &1.87    &\underline{4.44}    &4.09    &$ \bf 5.45^*$    &22.75\% \\
                            & NDCG@20 &0.91    &0.95    &1.20    &1.12    &\underline{1.69}    &1.03    &1.09    &1.17    &1.58    &1.37    &$ \bf 2.37^*$    &40.24\%  \\
                            \midrule 
\multirow{4}{*}{Yelp}       & Prec@10 &1.45    &1.95    &3.79    &2.94    &2.22    &1.51    &1.89    &1.80    &\underline{4.54}    &4.18    &$ \bf 5.63^*$    &24.01\% \\
                            & NDCG@10 &0.82    &1.26    &1.84    &1.54    &1.78    &1.02    &1.22    &1.11    &\underline{2.15}    &1.76    &$ \bf 3.09^*$    &43.72\% \\
                            & Prec@20 &2.17    &2.65    &4.95    &3.71    &3.65    &2.29    &2.62    &2.53    &\underline{7.94}    &7.59    &$ \bf 9.16^*$    &15.37\% \\
                            & NDCG@20 &1.30    &1.56    &2.40    &1.73    &1.93    &1.28    &1.60    &1.30    &\underline{3.02}    &2.51    &$ \bf 3.69^*$    &22.19\% \\
                            \bottomrule
\end{tabular}
\label{performance}
\end{table*}

It is worth noting that the Prec@$k$ metric does not take into account the ranking of an item, as long as it appears within the top-$k$ recommendations. Conversely, NDCG@$k$ emphasizes the item rank, which is crucial in scenarios where the order of recommendations matters. 
% Additionally, for both Prec@$k$ and NDCG@$k$, higher metric values signify better performance.  
In this study, we report the results with $k$ = 10 and 20.

\subsection{Baselines}

To evaluate the performance of the proposed \baby, we use following ten competitive methods in this study as baselines:
\begin{itemize}
    % \item \textbf{S-POP} recommends the most frequent items in the current session.  
    
    \item \textbf{SKNN} recommends items based on the similarity between the current session and the other sessions. 
     
    \item \textbf{NARM}~\cite{NARM} utilizes RNNs with an attention mechanism to capture the user’s main purpose. 
    
    \item \textbf{SASRec}~\cite{SASRec} employs the self-attention mechanism to capture sequential patterns of user behaviors. 
    
    \item \textbf{SR-GNN}~\cite{SR-GNN} builds session graphs and applies graph neural networks to capture item transitions.

    \item \textbf{RNS}~\cite{RNS} is a representative method incorporating reviews into the task to represent users and items.

    \item \textbf{ICLRec}~\cite{chen@WWW2022} mines user latent intents with contrastive learning for enhancing user behavior understanding.

    \item \textbf{UniSRec}~\cite{hou@KDD2022} represents items with description text for  obtaining universal sequence representations.

    \item \textbf{A-Mixer}~\cite{Zhang@WSDM2023} (\ie Atten-Mixer) leverages multi-level user intents to achieve multi-level reasoning on item transitions.
    
    \item \textbf{MCLRec}~\cite{Qin@SIGIR2023} presents meta-optimized contrastive learning to achieve informative data and model augmentations.
    
    \item \textbf{ACTSR}~\cite{Zhou@CIKM2023} enhances Transformer-based methods by calibrating unreliable attention weights in transformer layers.

\end{itemize}

\subsection{Implementation Details}

We employ ChatGPT-3.5\footnote{{https://chat.openai.com}} to conduct attribute-opinion extraction. Based on attribute popularity, the attributes utilized in each dataset are outlined in Table~\ref{attribute}. To strike a balance between efficiency and effectiveness, we focus on seven most popular attributes for each dataset. In order to mitigate the randomness of LLM, we execute the opinion extraction process five times, considering the union of these outputs as the final attribute-opinion pairs. 

To ensure a fair comparison, the hyper-parameters of \baby and all baselines are determined via grid search according to their performance on Prec@20 in the validation set. In our \baby, we investigate the dimension of attribute-specific user/item embedding $d$ in \{4, 8, 16, 32\}.
% and the number of attributes $N$ in \{ 5, 6, 7, 8, 9\}. 
As in~\cite{RNS}, the averaged embeddings of last 5 items ($l$=5) is used to represent a user's recent interest.
We set the mini-batch size to 512 and use the Adam optimizer with an initial learning rate of 0.001 to optimize \baby. We have released all source codes for broader accessibility and reproducibility\footnote{\url{https://github.com/Zhang-xiaokun/FineRec}}.
% , including datasets, prepocessing codes and model codes,  

\section{Results and Analysis}

\subsection{Overall Performance (RQ1)}\label{sec:overall}

The performance of all baselines and \baby on four datasets is detailed in Table~\ref{performance}, where we can obtain the following insights:

Firstly, the performance of baselines varies greatly across different datasets. Taking the ACTSR as an example, it shows best results among all baselines in Prec metric on Cellphones, while obtaining inferior performance in other contexts. Similarly, SASRec excels on Beauty dataset but performs poorly on others. These facts signify the complexity and difficulty of the sequential recommendation task, which requires capturing users' evolving interests from their historical behaviors. Moreover, most existing methods handle the task in an implicit way, primarily focusing on users' implicit clicking behaviors. Such a manner, unfortunately, often fails to grasp user genuine intents, thereby significantly limiting their effectiveness. 

Secondly, RNS, as a representative work incorporating reviews into SR, exhibits competitive performance in certain scenarios, \ie with the NDCG metric in Cellphones and Sports. Instead of modeling implicit user actions, RNS explicitly mines user interests from their reviews, enhancing its grasp of user intents. Nevertheless, its performance is less satisfactory on the Beauty and Yelp datasets. The primary drawback of RNS lies in its coarse paradigm for modeling a review as a whole. Indiscriminately blending distinct attitudes towards various attributes within a single review, RNS struggles to capture a user's  specific preferences for each attribute. This leads to its inconsistent performance across different datasets.

\begin{table*}[t]
% \small
\tabcolsep 0.05in 
  \centering
    \caption{The effect of diversity-aware convolution operation.}
    \begin{tabular}{c cc cc cc cc}  
    \toprule  
    \multirow{2}*{Method}& 
    \multicolumn{2}{c}{Cellphones}&\multicolumn{2}{c}{Beauty}&\multicolumn{2}{c}{Sports}&\multicolumn{2}{c}{Yelp}\cr  
    \cmidrule(lr){2-3} \cmidrule(lr){4-5} \cmidrule(lr){6-7} \cmidrule(lr){8-9}
    &Prec@20&NDCG@20 &Prec@20&NDCG@20 &Prec@20&NDCG@20 &Prec@20&NDCG@20\cr  
    \midrule  
    % POP&6.71&1.65&1.33&0.30&0.89&0.20\cr  
    % SR-GNN      &25.14&15.88    &42.74&35.01    &34.29&27.66\cr
    Best baselines  &8.84    &3.79    &7.93    &2.80    &4.44    &1.69    &7.94    &3.02\cr  
    w/o diver    &11.04    &4.29    &8.23    &3.18    &5.05    &2.04    &8.73    &3.47\cr
	{\bf \baby} &{ $ \bf 11.89^*$ }&{$\bf 4.75^*$} 
	            &{$\bf 9.25^*$ }&{ $ \bf 3.77^*$ }
	            &{$\bf 5.45^*$}&{$\bf 2.37^*$ }
                &{$\bf 9.16^*$}&{$\bf 3.69^*$ }
             % &{$\bf 19.33^*$}&{$\bf 4.76^*$ }
             \cr
	\bottomrule
    \end{tabular}

    % \end{threeparttable}  
    \label{diversity}
\end{table*}

Thirdly, the most recent approaches, namely MCLRec and ACTSR, generally outperform other baselines. Their impressive performance could be largely attributed to their utilization of cutting-edge techniques. Specifically, MCLRec leverages meta-optimized contrastive learning to enrich user behavior data. As to ACTSR, it calibrates attention weights within Transformer architectures to better adapt to the task. However, despite equipping with advanced techniques, these methods are unable to obtain consistent performance across various contexts. This once again suggests the limitations of implicit manner for modeling user behaviors in the task of sequential recommendation.

\begin{figure}[t]
  \centering
  \includegraphics[width=0.95\linewidth]{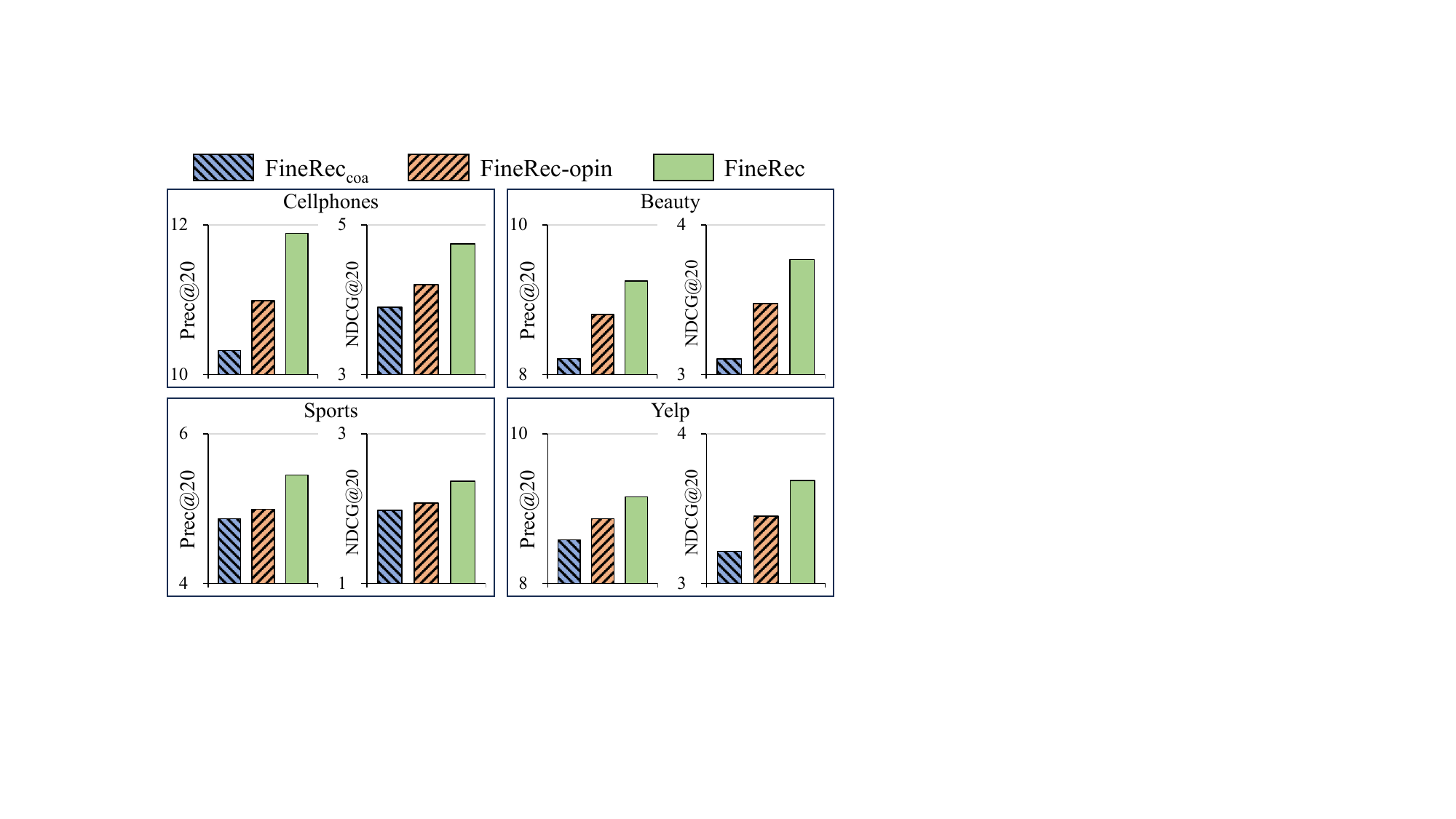}
  \caption{The effect of fine-grained manner on SR.}\label{fineEffect}
%  \Description{}
\end{figure}

Finally, the proposed \baby achieves consistent improvements over all baselines in terms of all metrics on all datasets, which demonstrates its effectiveness for sequential recommendation. In particular, \baby outperforms the best baselines in Prec@20 and NDCG@20 by 34.50\% and 25.33\% on Cellphones, 16.65\% and 34.64\% on Beauty, 22.75\% and 40.24\% on Sports, and 15.37\% and 22.19\% on Yelp. We believe that the consistent superiority of \baby over current start-of-the-art methods comes from its fine-grained manner in handling the task. Benefiting from finely representing users and items via attribute-opinions, \baby can identify fine-grained user preferences and item characteristics on various attributes. This manner enhances the prediction accuracy, significantly contributing to the overall effectiveness of the proposed \baby.

\subsection{The effect of fine-grained manner for handling SR (RQ2)}\label{sec:fine}

The key innovation of our \baby lies in its fine-grained manner in handling SR, achieved via exploring attribute-opinions from user-item reviews. In order to validate the effectiveness of this novel paradigm, the following variants of \baby are designed: ``\babyx$_{coa}$'' builds a single user-review-item graph based on the entire user-item interactions. In this graph, users and items form the nodes, and their reviews serve as the edges. That is, \babyx$_{coa}$ coarsely models a review as a whole, without distinguishing users' various opinions on different attributes. Besides, ``\babyx-opin'' creates an attribute-specific user-item graph for each attribute, without considering the related opinions between users and items. It relies on the conventional GCN to update user and item embeddings under each attribute. 

From~\fig~\ref{fineEffect}, we can obtain the following key points: (1) \babyx-opin achieves better performance than \babyx$_{coa}$. It underscores the rationality of building distinct attribute-specific sub-graphs for various attributes. This distinction aligns with the fact that users exhibit varied preferences and items display different characteristics across various attributes. By recognizing these differences, \babyx-opin obtains improvements over \babyx$_{coa}$; (2) \babyx-opin is defeated by \baby with a large margin, which highlights the significance of jointly exploring attributes and opinions. For an attribute, users may express completely different attitudes. Thus, only considering attributes while ignoring corresponding opinions, \babyx-opin fails to make accurate predictions; and (3) The \baby surpasses both its variants in all cases, validating the efficacy of our fine-grained manner in handling the task. By delving into the attribute-opinions within reviews, \baby is able to uncover user preferences and item characteristics, significantly improving the recommendation performance. 

\subsection{The effect of diversity-aware convolution operation (RQ3)}\label{sec:diver}

To tackle the opinion diversity issue in fine-grained user and item representation learning, we introduce a diversity-aware convolution operation. The variant "w/o diver" discards the diversity-aware convolution operation. Instead, it employs a straightforward approach by summing all adjacent opinion embeddings and item/user embeddings to update the user/item embeddings.  That is, it omits the diversity of opinions. Additionally, we include the results of the best-performing baselines for each metric to provide a comprehensive comparison.

Table~\ref{diversity} clearly shows that \baby consistently outperforms the "w/o diver" variant in all scenarios. It demonstrates the effectiveness of our specially devised diversity-aware convolution operation. By focusing on the diversity of opinions in user/item representation learning, this operation achieves two critical objectives: (1) grasping user varying opinions on distinct items, and (2) providing comprehensive portrayal of item characteristics based on various user opinions. Consequently, this operation enables \baby to obtain robust user/item representations under each attribute. In addition, the underperformance of the best baselines compared to the "w/o diver" variant further indicates the efficiency of our fine-grained manner in tackling sequential recommendation.

\subsection{The effect of interaction-driven fusion mechanism (RQ3)}\label{sec:inter}

\begin{figure}[t]
  \centering
  \includegraphics[width=0.95\linewidth]{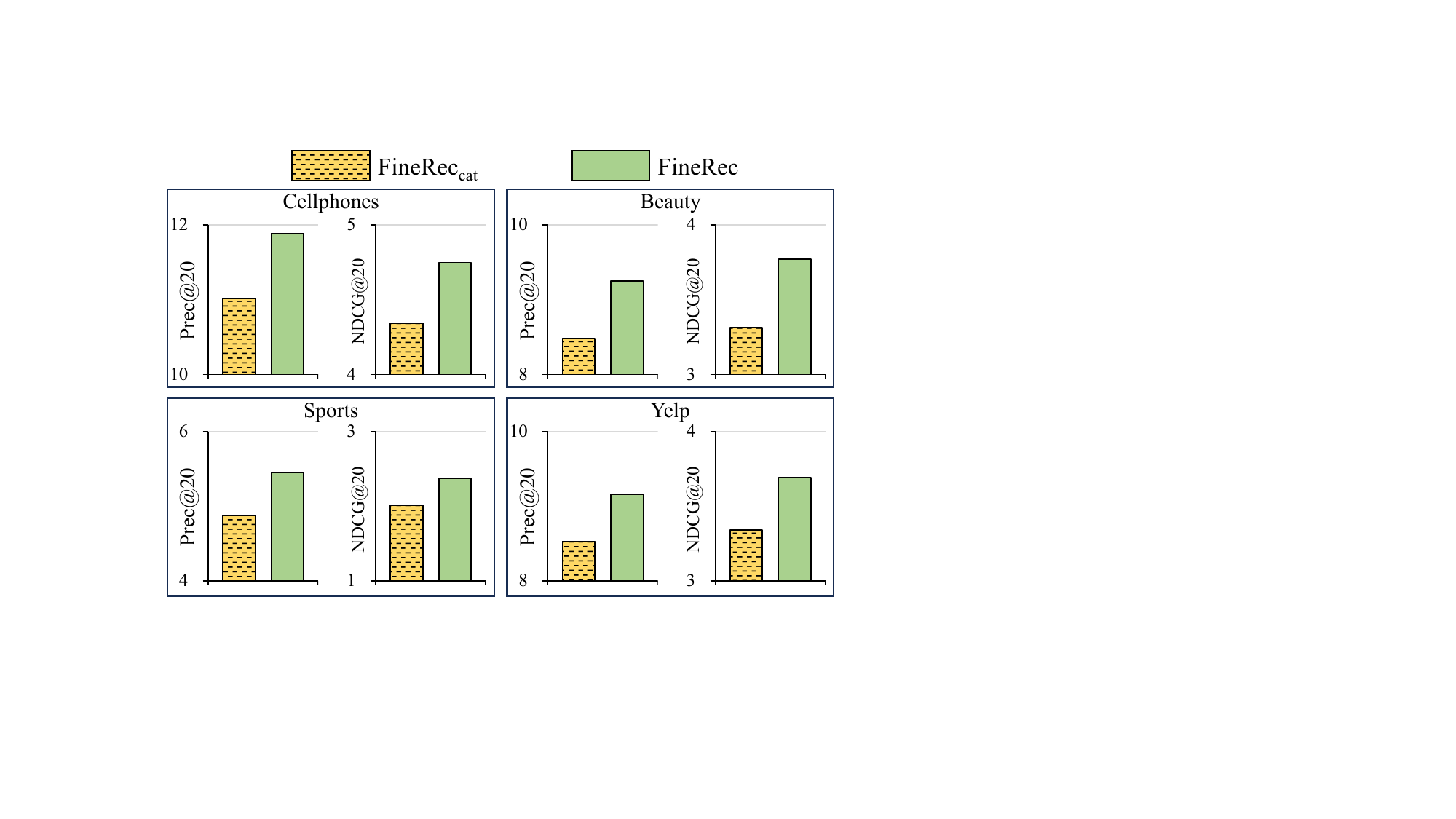}
  \caption{The effect of interaction-driven fusion mechanism.}\label{interfuse}
%  \Description{}
\end{figure}

To comprehensively understand user behaviors, we present an interaction-driven fusion mechanism that integrates user/item representations across all attributes. This mechanism employs informative user-item interactions to guide the fusion of attribute-specific user/item embeddings. In contrast, the \babyx$_{cat}$ variant employs a conventional approach for the fusion. It concatenates the embeddings from different attributes and then merges them using a Multi-Layer Perception (MLP), bypassing our specialized fusion mechanism.

As can be observed from~\fig~\ref{interfuse}, \baby achieves better performance over \babyx$_{cat}$. It validates the effectiveness of our interaction-driven fusion mechanism in integrating attribute-specific embeddings. We believe that the user-item interactions reflect the relationships, notably the similarity at the embedding-level, between users and items. By harnessing these similarities, the designed fusion mechanism gains a deeper understanding of how various attributes influence user behaviors. Therefore, the proposed mechanism facilitates an effective fusion process, thereby establishing the superiority of \baby over \babyx$_{cat}$.

\subsection{Hyper-parameter Study (RQ4)}\label{sec:hyper}

\begin{figure}[t]
  \centering
  \includegraphics[width=0.98\linewidth]{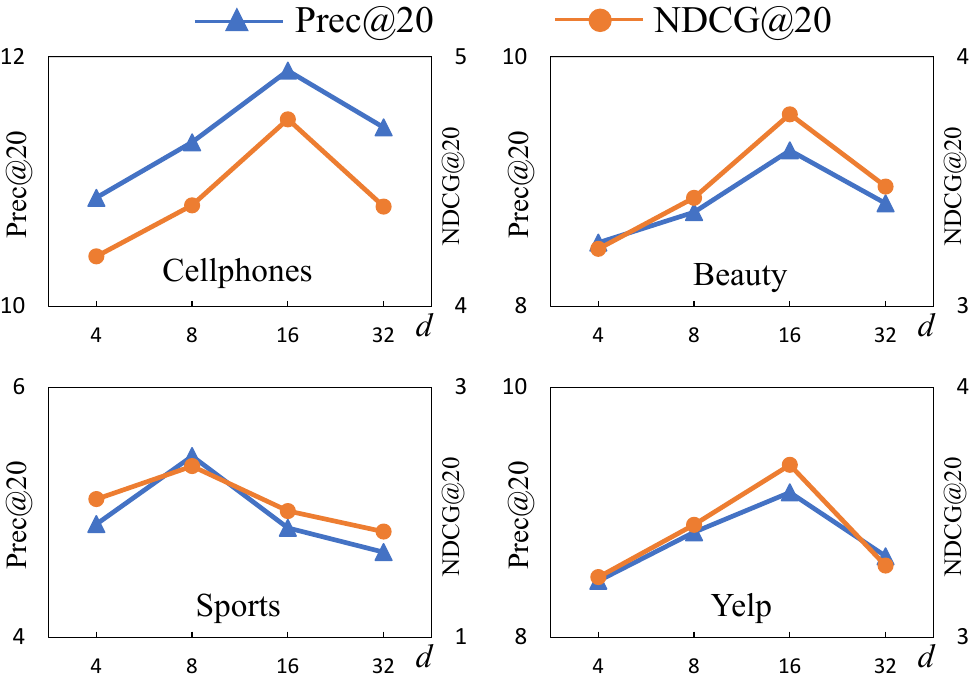}
  \caption{The impact of hyper-parameter $d$.}\label{hyper}
%  \Description{}
\end{figure}

In this section, we investigate the influence of the main hyper-parameter $d$ on \baby's performance. Parameter $d$ denotes the dimension of attribute embeddings, opinion embeddings, as well as attribute-specific user/item embeddings. The performance trends of \baby across various $d$ values, evaluated using Prec@20 and NDCG@20 metrics on all datasets, are illustrated in Figure~\ref{hyper}. The performance curve reveals that as $d$ increases, \baby initially shows improvements but eventually experiences a decline. This pattern suggests that while a larger $d$ enhances representation capability, leading to better performance, an excessively high $d$ might cause overfitting, thereby degrading performance. Interestingly, we find that a relatively small $d$ value suffices for optimal performance in \baby, with $d$ set to 8 for Sports, and 16 for Cellphones, Beauty, as well as Yelp datasets. Our \baby aims to capture specific user/item preferences/characteristics on each attribute, instead of indiscriminately modeling them as a whole like existing methods. Benefiting from such a fine-grained manner, \baby can accurately represent users or items under each attribute at a low cost (\ie a small number of embedding dimension). It is both efficient and effective. We believe that this merit of \baby significantly enhances its practical applicability.

\subsection{Case Study (RQ5)}\label{sec:case}

\begin{figure*}[t]
  \centering
  \includegraphics[width=0.95\linewidth]{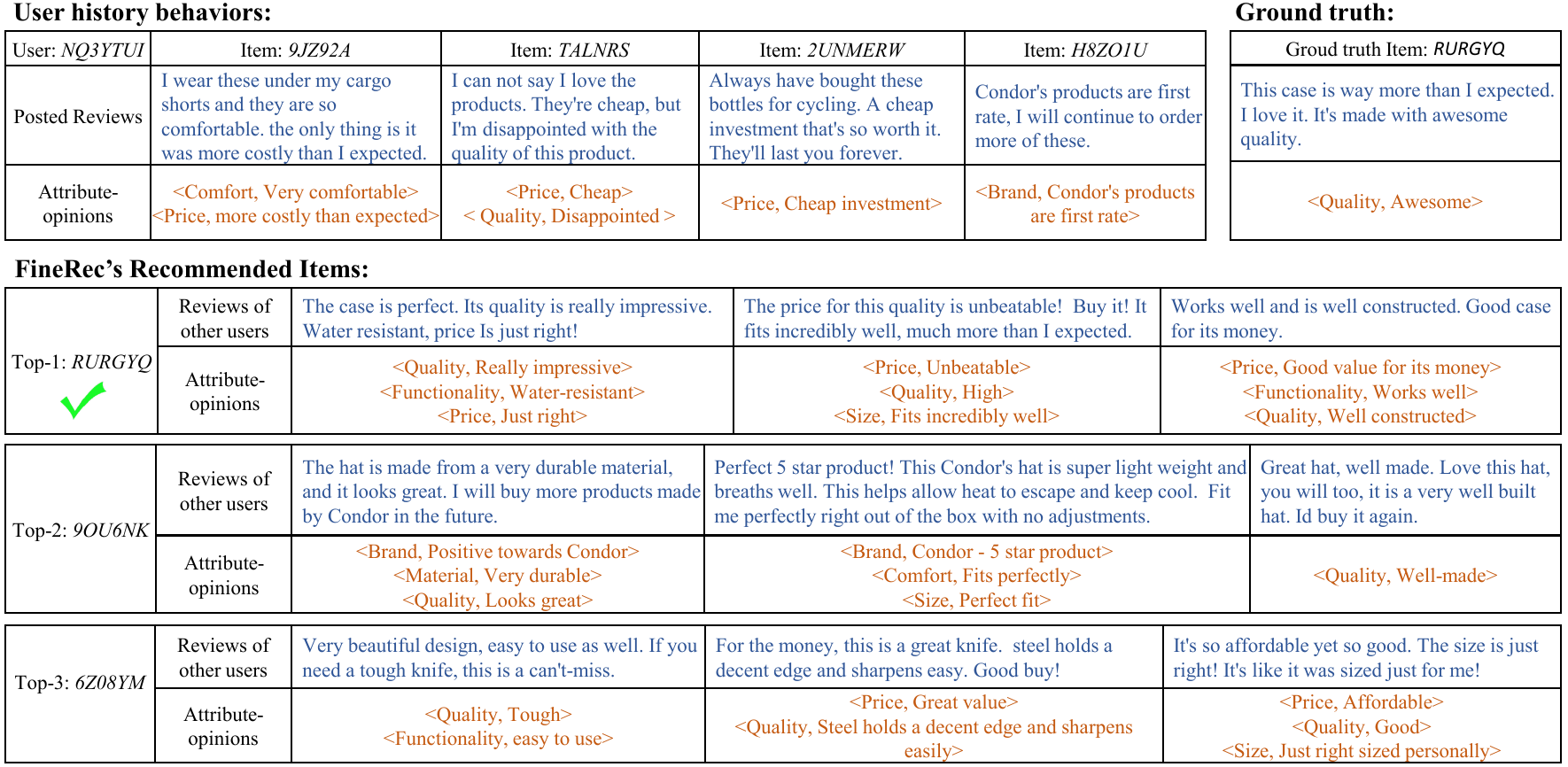}
  \caption{A case from the Sports dataset. We present user-item reviews (Blue), corresponding attribute-opinions (Orange), and the recommended items of \baby with other users' reviews.}\label{case}
%  \Description{}
\end{figure*}

In order to intuitively scrutinize the effects of \baby in generating recommendations, we randomly select an instance from the dataset Sports. This real-world case is presented in~\fig~\ref{case}, where we detail a user's historically interacted items, reviews he posted for each item, the attribute-opinion pairs extracted using LLM, and the ground truth item. Additionally, we display recommended items from \baby at the low part of the figure. For these recommended items, we include three reviews written by other users, along with their respective attribute-opinion pairs. Due to space constraints, we present the top-3 recommended items of \baby.

The following observations are noted in~\fig~\ref{case}: (1) The LLM-based attribute-opinion extraction can effectively derive attribute-opinion pairs from user-item reviews. As shown in the case, the extracted attribute-opinions are generally informative and accurate. We believe that this efficiency likely stems from the following facts: (a) the LLM’s extensive training on massive text data encapsulates it with rich language knowledge; and (b) our approach of directing the LLM towards specific attributes instead of from scratch to generate attribute-opinions alleviates its inherent "hallucinations". (2) Our proposed fine-grained manner does facilitate effective recommendations. As shown in the case, the attributes the user cares about are hit by our \baby. More specifically, the attributes mentioned by the user are mostly overlapped with those of items provided by \baby, \ie \{Price, Quality\}. This success is reasonable since that our \baby represents users and items with fine-grained attribute-opinions instead of conventional implicit ID number. More surprisingly, based on user specific attribute-opinions, our \baby can provide highly personalized recommendations. Concretely, the user expressed "disappointed" to the "Quality" of the item "TALNERS". Our \baby can correctly recommend ground truth item "RURGYQ" which has been rated by other users as of "really impressive", "high" and "well constructed" Quality. Evidently, such type of recommendation can greatly improve user satisfaction. (3) The proposed \baby can effectively discern fine-grained user preferences and item characteristics, offering relevant recommendations accordingly. For example, for the user who showed a preference for the "Condor" brand in his reviews, \baby is able to recommend an item associated with the same brand, where the item brand is identified from other reviews.

\section{Conclusion and Future Work}
In this study, we propose a novel framework \baby to explore fine-grained sequential recommendation via mining the attribute-opinions from reviews. 
% Three obstacles in the fine-grained modeling are identified as: (1) extraction of attribute-opinion ; (2) fine-grained user and item representation with diverse opinions; (3) recommendation from various user preferences and item characteristics in distinct attributes. To address these issues, an innovative framework \baby is proposed. 
Specifically, \baby employs a Large Language Model to extract attribute-opinions from reviews. For each attribute, we create an attribute-specific user-opinion-item graph, encoding fine-grained user preferences and item characteristics. Within these graphs, a diversity-aware convolution operation is devised to achieve fine-grained user and item representations, particularly handling the diversity of opinions. Lastly,an interaction-driven fusion mechanism is employed, harnessing user-item interactions to integrate attribute-specific embeddings for generating final recommendations. Comprehensive experiments on four real-world datasets highlight \baby's superiority over current state-of-the-art methods. Further analysis also verifies the effectiveness of our finely-grained manner for the task.

As to future work, our fine-grained approach opens up exciting avenues for enhancing the interpretability of recommendation tasks. The prospect of providing explanations rooted in the fine-grained user preferences and item characteristics is both attractive and promising. Such a perspective could offer a more fine-grained and personalized understanding of recommendations, significantly enhancing the user experience.
% we plan to optimize the extraction for attribute-opinions in our \baby. We are optimistic about the advancements in Large Language Models (LLMs), anticipating more robust and comprehensive systems that could facilitate direct extraction of attributes and their associated opinions. Additionally,

\section*{Acknowledgement}
This work has been supported by the Natural Science  Foundation of China (No.62076046, No.62376051, No.62366040). We would like to thank the anonymous reviewers for their valuable comments.

\newpage

\onecolumn
\begin{multicols}{2}
   \bibliographystyle{ACM-Reference-Format}
   \bibliography{ref}
\end{multicols}

% \bibliographystyle{ACM-Reference-Format}
% \bibliography{ref}

\end{document}